# Nano-scale Inhomogeneous Superconductivity in Fe(Te$_{1-x}$Se$_x$) Probed by Nanostructure-transport

Chunlei Yue[§†], Jin Hu[§†], Xue Liu[†], Zhiqiang Mao[†], and Jiang Wei*[†]

[†]Department of Physics and Engineering Physics, Tulane University, New Orleans, Louisiana 70118
[§]These authors contribute equally to this work.

*Address correspondence to jwei1@tulane.edu

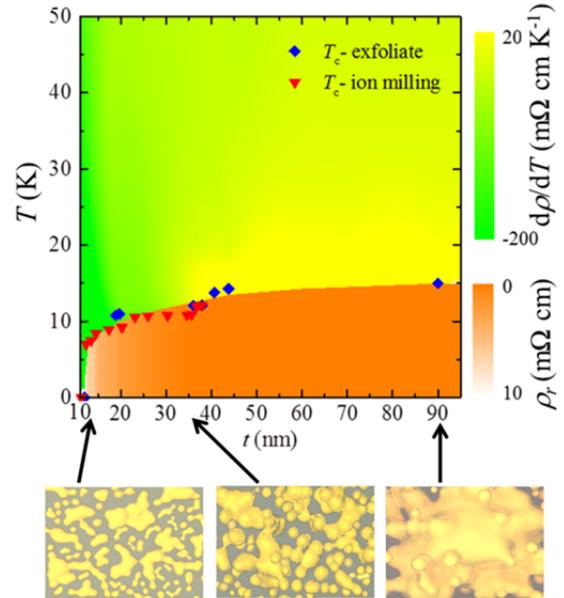

ABSTRACT: Among iron based superconductors, the layered iron chalcogenide Fe(Te$_{1-x}$Se$_x$) is structurally the simplest and has attracted considerable attentions. It has been speculated from bulk studies that nanoscale inhomogeneous superconductivity may inherently exist in this system. However, this has not been directly observed from nanoscale transport measurements. In this work, through simple micromechanical exfoliation and high precision low-energy ion milling thinning, we prepared Fe(Te$_{0.5}$Se$_{0.5}$) nano-flake with various thickness and systematically studied the correlation between the thickness and superconducting phase transition. Our result revealed a systematic evolution of superconducting transition with thickness. When the thickness of Fe(Te$_{0.5}$Se$_{0.5}$) flake is reduced down to 12nm, *i.e.* the characteristic length of Te/Se fluctuation, the superconducting current path and the metallicity of normal state in Fe(Te$_{0.5}$Se$_{0.5}$) atomic sheets is suppressed. This observation provides the first direct transport evidence for the nano-scale inhomogeneous nature of superconductivity in Fe(Te$_{1-x}$Se$_x$).

The discovery of layered iron pnictides [1-10] and iron chalcogenide [11-14] superconductors has provided a model system for the study of the interplay between high temperature superconductivity and magnetism. Iron chalcogenide Fe(Te$_{1-x}$Se$_x$) has the simplest crystal structure among all discovered Fe-based superconductors. This system displays a complex coexistence of two antiferromagnetic correlations with in-plane wave vector ($\pi$,0) [15, 16] and ($\pi$,$\pi$) [17, 18]. The relative strength of these two magnetic correlations can be tuned by changing the Te/Se ratio, which leads to an unusual evolution from ($\pi$,0) long-range magnetic order to superconductivity with ($\pi,\pi$) spin resonance [19, 20]. The ($\pi$,0) magnetic correlations are found to cause incoherent magnetic scattering, which is detrimental to superconductivity [20]. As a result, the superconductivity in Fe(Te$_{1-x}$Se$_x$) occurs only when the ($\pi$,0) magnetism is greatly suppressed by much higher concentration of Se, *e.g.*, $x>0.29$ [19, 20]. Although earlier experiments highlighted the "coexistence" of superconductivity and magnetic order in low Se concentration samples [21-23], it has been later clarified that the "superconductivity" observed in this region is only a trace, which cannot be probed by bulk property measurements such as specific heat [19, 20, 24]. On the other hand, nanoscale chemical inhomogeneity, *i.e.* non-uniform spatial distribution of Te/Se, is immanent for Fe(Te$_{1-x}$Se$_x$) crystals as evidenced by scanning transmission electron microscopy (STEM) and electron energy-loss spectroscopy (EELS) analyses[25, 26]. Such chemical inhomogeneity is believed to be the cause of the inhomogeneous superconductivity. That is, the nano-scale separation of compositions lead to the formation of small superconducting islands in some local areas with higher Se concentration and the percolation of such superconducting islands can result in a remarkable superconducting transition in electronic transport measurements [19, 20]. Yet, electrical transport measurement on nanoscale samples, which may directly connect the chemical inhomogeneity to the percolated phase separation in Fe(Te$_{1-x}$Se$_x$) crystals has not been reported. Bellingeri *et. al.* [27] reported a suppression of $T_c$ with the reduction of the thickness in Fe(Te$_{1-x}$Se$_x$) epitaxial films grown by pulsed laser deposition (PLD) technique. However, such a thickness dependence on $T_c$ is found to be largely associated with the strain effect from the substrate as pointed out by the authors [27]. Okazaki et al. found a difference in $T_c$ for three relatively underdoped x=0.35 thin flake samples[28]. Yet the inconsistent superconducting transition behavior among the flakes with different thickness indicates their bulk sample may be macroscopically nonuniform. In this work, to directly reveal the nano-scale inhomogeneous superconductivity in Fe(Te$_{1-x}$Se$_x$), we have performed detailed transport studies on strain-free Fe(Te$_{0.5}$Se$_{0.5}$) thin flakes of various thickness prepared by two approaches, i.e. microexfoliation and low-energy ion milling thinning. We have observed a systematic suppression of superconductivity with the decrease of flake thickness, which can be attributed to the nano-scale superconducting-normal phase separation. Such percolation-like transition is further verified and supported by our model simulations. These observations provide direct transport evidence for the nano-scale inhomogeneous superconductivity in Fe(Te$_{1-x}$Se$_x$) for the first time.

RESULTS AND DISSCUSSION

The crystal structure of Fe(Te$_{1-x}$Se$_x$) can be viewed as the stacking of sandwich-like Te/Se-Fe-Te/Se layers as shown in the inset of Figure 1(a). Within each layer the Fe atoms are tetrahedrally coordinated by Te/Se atoms, forming a charge neutral layer. Such structural characteristic of Fe(Te$_{1-x}$Se$_x$) is in sharp contrast to iron pnictide superconductors that are composed of alternating charged layers. In fact, the stacking of neutral sandwich layer of Fe(Te$_{1-x}$Se$_x$) is similar to that of transition metal dichalcogenides, such as MoS$_2$, which also consist of charge neutral chalcogen-metal-chalcogen sandwiches layers[29, 30]. The existence of weak van der Waals (vdW) bonding between adjacent charge neutral layers makes it easy to exfoliate Fe(Te$_{1-x}$Se$_x$) crystals down to atomically thin sheets, providing an excellent platform to explore its properties in nano-scale.

The optimally doped ($x$=0.5) bulk single crystals were synthesized using a flux method as reported elsewhere[31]. The structure and composition of bulk single crystals were examined using X-ray diffraction (XRD) and energy dispersive X-ray spectrometer (EDX), respectively. Our previous bulk measurements such as heat capacity [19, 20, 32] and neutron scattering [18] measurements has already revealed a bulk superconducting volume fraction > 90% at $T_c$~15K for $x$=0.5 crystal. The fact that the superconducting volume fraction does not reach 100% implies possible existence of inhomogeneous superconductivity to some extent. Atomically thin Fe(Te$_{0.5}$Se$_{0.5}$) crystals can be obtained on Si/SiO$_2$ substrate via microexfoliation (see Methods), which is widely recognized to be able to maintain the highest quality of crystal and has been extensively used for fabricating 2D crystals of graphene and transition metal dichalcogenides[33-36]. The quality and stability of the exfoliated Fe(Te$_{0.5}$Se$_{0.5}$) thin flakes were examined by using the transmission electron microscope (TEM) measurements. As shown in Figure 1(b), we have observed clear electron diffraction patterns on those as-exfoliated Fe(Te$_{0.5}$Se$_{0.5}$) flakes by using selected area electron diffraction (SAED). All those [001]-zone electron diffraction spots can be indexed according to the crystal structure of Fe(Te$_{0.5}$Se$_{0.5}$), suggesting the excellent crystallinity for the as-exfoliated thin flake.

The Fe(Te$_{0.5}$Se$_{0.5}$) nano-devices with Ohmic contact were fabricated using standard electron-beam lithography (see Methods). To eliminate the influence of contact resistance for the transport measurements, four-terminal configuration of the device has been adopted, as illustrated in Figure 1(a). In Figure 1(c) we plot the temperature dependence of resistivity for Fe(Te$_{0.5}$Se$_{0.5}$) with various thickness ($t =$ 12 - 90 nm) obtained by the exfoliation technique. For the relatively thick flake ~ 90 nm (red line), a sharp superconducting transition at $T_c$ ~15K is observed, with the normal state resistivity above $T_c$ being metallic-like, similar to the observations in bulk Fe(Te$_{0.5}$Se$_{0.5}$) samples [19, 20, 37]. With decreasing thickness, we have observed a systematic suppression of $T_c$, which is accompanied by the development of the non-metallic normal state and finite residual resistivity. Further reduction of thickness (below 18 nm) leads to the upturn of the residual resistivity, and this trend becomes more significant with the thickness approaching 12 nm. At $t = 12$ nm, superconductivity disappears and the sample displays non-metallic behavior over the whole temperature range.

Owing to the fact that the thickness of the thin flakes cannot be precisely controlled using the microexfoliation method, it is difficult to obtain a precise description of the evolution of superconductivity with thickness. To establish a completed thickness dependence of the electronic properties of Fe(Te$_{0.5}$Se$_{0.5}$) system, we have applied an alternative approach to produce thin flakes with desired thickness, i.e., using low-energy Argon ion milling to thin down Fe(Te$_{0.5}$Se$_{0.5}$) crystals to thin flakes. Ion milling is widely used to prepare thin specimen for TEM measurements [38]. The collimated ion beam bombards the surface of a specimen and reduces its thickness with precision of sub-nanometers. To minimize possible superficial damages to Fe(Te$_{0.5}$Se$_{0.5}$) thin crystal from ion bombardment, we choose low energy Argon (e.g. 600V) and small incident angle (5°) during the milling process. Given that both the Fe(Te$_{0.5}$Se$_{0.5}$) thin flake and the Si/SiO$_2$ substrate are milled conformally by ion bombardments, the thickness of the sample cannot be simply determined by measuring the height profile with AFM. Instead, we estimate the flake thickness by using a calibrated milling rate of 2.79 nm/minutes (see Methods). The surface of the Fe(Te$_{0.5}$Se$_{0.5}$) thin flakes checked by AFM revealed a unchanged roughness ~1nm before and after ion milling. This value is consistent with the roughness of Si$_3$N$_4$ substrate and implies that the exfoliated single crystal is atomically flat by contouring the substrate surface (see Figure 2(d)). Therefore, to exam any possible crystal damage produced by ion milling, we performed TEM study on an identical thin flake (initial thickness~27nm) before and after ion milling. As shown in Figure 2(a)-(c), after two successive 90 seconds of ion milling, the selected area electron diffraction (SEAD) pattern remains sharp and clear, without any additional interfering or distortion patterns being introduced, implying that the low-energy ion milling hardly causes any noticeable damage to Fe(Te$_{0.5}$Se$_{0.5}$) crystals.

We selected a device fabricated with an exfoliated thin flake with thickness of 37.86 nm ($T_c$ ~ 12K before ion milling) and then reduced its thickness by using ion milling as described above. The temperature dependence of resistivity was measured right after each step of thinning. As shown in Figure 2(e), similar to the results obtained on exfoliated flakes shown in Figure 1(c), the decrease of thickness by ion milling also leads to the development of non-metallicity and the suppression of superconductivity. We have summarized the thickness dependence of superconductivity of Fe(Te$_{0.5}$Se$_{0.5}$) thin flakes in the contour plot in Figure 2(f). It can be readily seen that the thickness dependence of $T_c$ shares a similar trend for exfoliated and ion milled thin flakes. A critical thickness $t_c$~ 12 nm where superconductivity completely disappears can be determined.

What is the origin of the suppression of superconductivity with reducing thickness in our Fe(Te$_{0.5}$Se$_{0.5}$) thin flakes? Although our observation of superconductivity suppression and non-metallic normal state in thin flake samples appears to be similar to the behaviors previously observed on epitaxial

Fe(Te$_{0.5}$Se$_{0.5}$) films [27], the lattice deformation, which is believed to play a major role in suppressing superconductivity in those epitaxial thin film samples [27], should not be the cause in our strain-free samples. Sample degrading could also lead to the absence of superconductivity, but our TEM observations have demonstrated the excellent crystallinity of thin flakes obtained from both thinning techniques. In addition, the suppression of superconducting transition is also commonly seen when the sample size is reduced below than the coherence length [39, 40]. However, given that both the in-plane and out-of-plane superconducting coherence length estimated from the upper critical field ($\xi_{//}$, $\xi_{\perp}$ ~ 3 nm) [41-44] for Fe(Te$_{1-x}$Se$_x$) are well below the critical thickness (~12nm), the dimensionality crossover is not likely the cause. It has also been reported that atomic disorder can lead to the suppression of superconductivity in homogeneous films through the combined effect of the electron interaction and impurity scattering [45]. Such a scenario is possible given the presence of disorder induced by Se substitution for Te in Fe(Te$_{1-x}$Se$_x$). Nevertheless, the situation may be different in Fe(Te$_{1-x}$Se$_x$) as compared to the polycrystalline metal films reported in previous studies (see ref. 45 and references therein). As we show below, our observations in strain-free samples should be attributed to nano-scale inhomogeneous superconductivity in Fe(Te$_{1-x}$Se$_x$) system.

To understand the suppression of superconductivity in thin flakes, we studied the normal state electronic properties of these devices, and summarized the derivative of the temperature dependence of the normal state resistivity d$\rho$/d$T$ in the contour plot presented in Figure 2(f). In the same contour plot we also include the residual resistivity $\rho_r$ at 2K, as well as the critical temperature $T_c$, to characterize the thickness dependence of superconductivity. Interestingly, we can readily find that the suppression of superconductivity in Fe(Te$_{0.5}$Se$_{0.5}$) devices concomitants with the loss of the normal state metallicity. As shown in the contour plot in Figure 2(f), in the thinner devices, the normal state metallicity weakens as shown by the more negative d$\rho$/d$T$, while the superconductivity is also suppressed as characterized by the lowered $T_c$ and enhanced $\rho_r$. Such signatures in our Fe(Te$_{0.5}$Se$_{0.5}$) devices reminisce the observations in bulk Fe(Te$_{1-x}$Se$_x$), where the bulk superconductivity occurs only when the normal state becomes metallic for $x > 0.3$ [20]. These similar crossovers for normal state properties and superconductivity suggest possible relevance in their origins.

As pointed out in Ref [20], when the long-range ($\pi$.0) antiferromagnetic order in FeTe is suppressed by increasing Se content $x$ above 0.1, the remnant short range ($\pi$.0) magnetic fluctuations leads to incoherent scatterings to charge carriers, which leads to localized electronic states and superconductivity suppression in low Se concentration samples ($x < 0.3$). With further Se substitution ($x > 0.3$), such incoherent scatterings is significantly suppressed, thus the normal state transport becomes metallic and bulk superconductivity occurs. Such a scenario, established on the variation of Se content in bulk materials, may also be applied to our Fe(Te$_{0.5}$Se$_{0.5}$) nano-flakes. As revealed in STEM and EELS analyses [25], the concentration of Te shows ~20% fluctuations from the average local composition, leading to nano-scale phase separation. As stated above, the regions with high Te content (*i.e.*, low Se) are expected to exhibit charge carrier localization due to strong incoherent magnetic scattering. However, in three dimensional (3D) case the superconducting paths are easy to form in such non-superconducting – superconducting percolation network, leading to good superconductivity properties in bulk materials. In contrast, in a nano flake with its size close to that of the minor non-superconducting phase, the percolation threshold that enables the connection of superconducting islands becomes much higher, as predicted by the 2D percolation theory. Such finite size effect can be illustrated in our simulation for the formation of superconducting paths in flakes with different thickness, which was performed based on the assumption of the presence of the ~10% non-superconducting phase, since the superconducting volume fraction estimated from specific heat measurements is ~ 90-92% for our optimally doped (x~0.4-0.5) single crystals[20, 24, 32]. As presented in Figure 3(a) and (b), the superconducting islands are well connected and form robust superconducting paths in very thick (3D) flakes. Therefore the electrons can detour the non-superconducting phase during transport measurements and a sharp zero-resistance transition can occur in those thick flakes (e.g., $t > 40$ nm devices shown in Figure 1(c)). The reduction of sample thickness maintains the lateral connections of these superconducting islands, but significantly restricts their vertical connections. Therefore, the formation of superconducting channels connecting both ends of the sample is limited as compared with thick flakes, as shown in Figure 3(c) and (d). Further decreasing flake thickness to a value comparable to the characteristic length of non-superconducting minor phase results the intersection of superconducting channel by the minor non-superconducting phase, as illustrated in Figure 3(e) and (f). Therefore, even individual superconducting islands still exhibit zero resistance below $T_c$, the resistance of the whole sample cannot reach zero due to the lack of continuous superconducting path throughout the sample. Under this circumstance, we can observe the development of the finite residual resistivity, as seen in Figure 1(c) and 2(e). Such a scenario is consistent with the observation in our 12.2 nm thick device (Figure 1(c)), which displays clear coexistence of the resistivity drop from the isolated superconducting islands, and the significant residual resistivity upturn that follows similar localization behavior of the non-superconducting Te-rich phase [20]. Indeed, the critical thickness for which superconductivity disappears is estimated to be ~12 nm in our Fe(Te$_{0.5}$Se$_{0.5}$) devices as shown in Figure 1(c), comparable to the scale of the Te content deviation (~ 10 nm) obtained in EELS measurements on a sample with similar composition, Fe(Te$_{0.55}$Se$_{0.45}$) [25]. It is worth noting that the superconducting volume fraction estimated from the bulk single crystals is in fact much greater than the expected percolation threshold for both 3D and 2D systems, implying the existence of the continuous superconducting path despite of the sample thickness. However, given that the percolation threshold differs for various types of lattices, the true value may be largely different from the theoretical value due to the complicate shapes of each superconducting island in our system. Nevertheless, the trend of the superconductivity suppression is in principle consistent with the percolation theory.

In addition to the Te/Se ratio discussed above, superconductivity in Fe(Te$_{1-x}$Se$_x$) system can also be controlled by excess Fe at the interstitial site of Te/Se layer [19, 20, 31]. That is because that the excess Fe enhances the ($\pi$, 0) magnetic fluctuations and induces incoherent magnetic scatterings, leading to charge carrier localization and superconductivity suppression.[19, 20] As pointed out by earlier studies, excess Fe stabilizes the crystal structure of Fe(Te$_{1-x}$Se$_x$) and inevitable during sample synthesis, despite that the nominal composition of the starting materials is stoichiometric [24, 37, 48]. Therefore, even in optimally doped samples (x>0.4) where the level of excess Fe is minimized [24, 37, 48], the presence of a very tiny amount of excess Fe leads to more inhomogeneous superconductivity in the system, and is expected to causes the suppression of superconductivity in nano-scale samples in a fashion similar to what have been discussed above.

Another approach to examine the presence of superconducting islands is magnetotransport measurement. The suppression of superconductivity by magnetic field can lead to significant magnetoresistivity (MR). We first checked such a scenario in a thick device ($t$ = 39 nm) with the superconducting transition occurring in 12-14K (Figure 1(c) and (d)). We have performed systematic MR measurements at various temperatures and summarized the normalized MR (=[$\rho(B)$-$\rho(B=0)$]/$\rho(B=0)$) in the contour plot in Figure 4(a). Above the onset $T_c$ (i.e. >14K), MR follows a very weak $B^2$ dependence (e.g. an example at 15K is shown in Figure 4(b)), suggesting the orbit MR due to Lorentz effect. Because Fe(Te$_{1-x}$Se$_x$) possesses very high upper critical field (> 40T) [41-44], a field of a few tesla can only slightly suppress $T_c$. Therefore MR becomes significant slightly below the onset $T_c$, but vanishes deep in the superconducting state due to the persistence of zero resistivity. As shown in Figure 4(b), for example, superconductivity at $T$ = 12.5 K is rapidly suppressed by field and resistivity tends to saturate to the normal state value, leading to a *superlinear* field dependence of MR. With decreasing temperature, the field become less efficient in suppressing superconductivity and MR at 8K is essentially zero. Similar MR enhancement near $T_c$ has also been observed in a series of Fe(Te$_{1-x}$Se$_x$) flakes with various thickness down to 12.2 nm (Figure 4(c) and (d)), indicating the presence of superconducting phase in these samples. Additionally, with the thickness approaching the critical thickness (~12 nm), we observed considerable non-zero MR well below $T_c$ (~4% at 8T and 5.5K) that follows $B^2$ dependence, suggesting that it should primarily arise from the non-superconducting island. Such observation, together with the low temperature residual resistivity upturn (see Figure 1(c)), clearly implies well separated superconducting islands by non-superconducting phase in this sample.

Finally, we should discuss the possibility of 2D superconductivity, which is described by the Berezinsky–Kosterlitz–Thouless (BKT) model characterized by a phase transition to a 2D topological order below a BKT temperature ($T_{BKT}$) [49, 50]. We have fitted the resistivity using the BKT model with the consideration of inhomogeneity and finite size effect [51]. As shown in Figure 1(d), the BKT model can almost describe the superconducting transition, suggesting the possible 2D superconductivity. However, as discussed above, the thicknesses of all the superconducting samples (*i.e.* >12nm) are well above the superconducting coherence length of Fe(Te$_{1-x}$Se$_x$) ($\xi_{//}$, $\xi_\perp$~3 nm [41-44]), making the 2D superconductivity less likely. More experiments such as *I-V* isotherms are necessary to exam the BKT signatures, which will be the scope of our follow up work.

CONCLUSIONS

In summary, through transport measurements on Fe(Te$_{0.5}$Se$_{0.5}$) nano-flake, we found that superconductivity is suppressed when the sample thickness is reduced below 12nm by microexfoliation or low-energy ion milling. Our analyses indicate that such superconductivity suppression is caused by the nano-scale phase separation owing to inhomogeneous distribution of Te/Se and/or the present of tiny amount of excess Fe. This observation provides the first direct transport evidence for the nano scale inhomogeneous superconductivity nature in iron chalcogenide system.

METHODS

Single-crystal thin flakes of Fe(Te$_{0.5}$Se$_{0.5}$) were achieved by using micro-mechanical exfoliation method. The fresh inner surface of Fe(Te$_{0.5}$Se$_{0.5}$) crystal was split open with a Scotch tape and then transferred to SiO$_2$/Si (10nm/290nm) wafer. The flake thickness was first estimated according to its color contrast under an optical microscope and then precisely measured by an atomic force microscope (AFM). Thin layers with thicknesses ranging from a few nanometers up to hundreds of nanometers can be produced.

To fabricate Fe(Te$_{0.5}$Se$_{0.5}$) nanodevice, four-terminal contact leads pattern was first created by standard electron beam lithography, then followed by 5nm Ti and 50 nm Au deposition as the contact metal with an e-beam evaporator. Ohmic contact can be realized on such device.

For TEM study, thin Fe(Te$_{0.5}$Se$_{0.5}$) flakes were first exfoliated onto a silicon substrate with a few nanometer poly acrylic acid (PVA) layer and 300nm polymethyl-methacrylate (PMMA) coating as top layer. PMMA layer with attached flakes was then released from the substrate by dissolving PVA layer in water and further transferred onto a rigid silicon based TEM grid with 30nm thick Si$_3$N$_4$ membrane window.

The low-energy ion milling on as-exfoliated Fe(Te$_{0.5}$Se$_{0.5}$) flakes was carried out on an AJA-IM ion milling system with milling parameters set as beam voltage/acceleration voltage of 600V/120V and ion beam incident angle of 5°. To calibrate the milling rate, electron-beam lithography was used to pattern a layer of PMMA that covers half of the Fe(Te$_{0.5}$Se$_{0.5}$) crystal and exposes the other half to ion milling. After ion milling thinning, the PMMA was then dissolved in Acetone and the resulted thickness difference on the same crystal between milled and un-milled area was measured by AFM to obtain a milling rate.

*Conflict of Interest:* The authors declare no competing financial interest.

*Acknowledgment.* This work is supported by the DOE under grant DE-SC0014208. We acknowledge Coordinated Instrument Facility (CIF) of Tulane University for the support of Transmission Electron Microscope (TEM) and Dr. Jibao He for the discussion of TEM data.

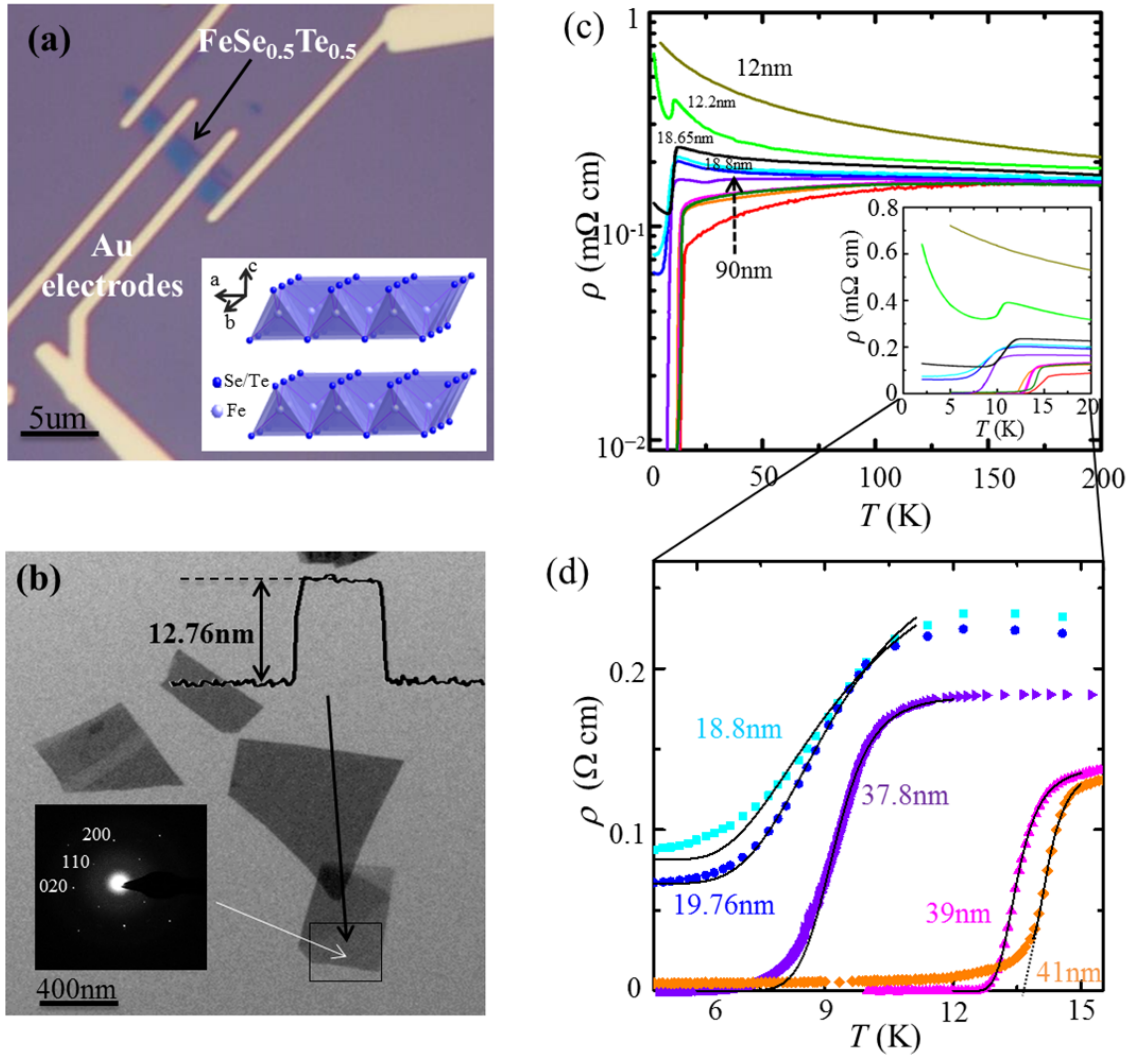

Figure 1. (a) Optical image of a Fe(Te$_{0.5}$Se$_{0.5}$) nano device. The inset shows the crystal structure of Fe(Te$_{1-x}$Se$_x$). (b) TEM image of the Fe(Te$_{0.5}$Se$_{0.5}$) thin flakes. Lower inset: SAED diffraction pattern of the Fe(Te$_{0.5}$Se$_{0.5}$) thin flake. Upper inset: AFM height profile of the thin flake used for SAED diffraction. (c) Temperature dependence of resistivity of the Fe(Te$_{0.5}$Se$_{0.5}$) nanodevices obtained using microexfoliation method. Inset: low temperature (< 20K) resistivity of these nanodevices. (d) Fitting of the superconducting transition using the Berezinsky–Kosterlitz–Thouless model with the consideration of inhomogeneous and finite size effect.

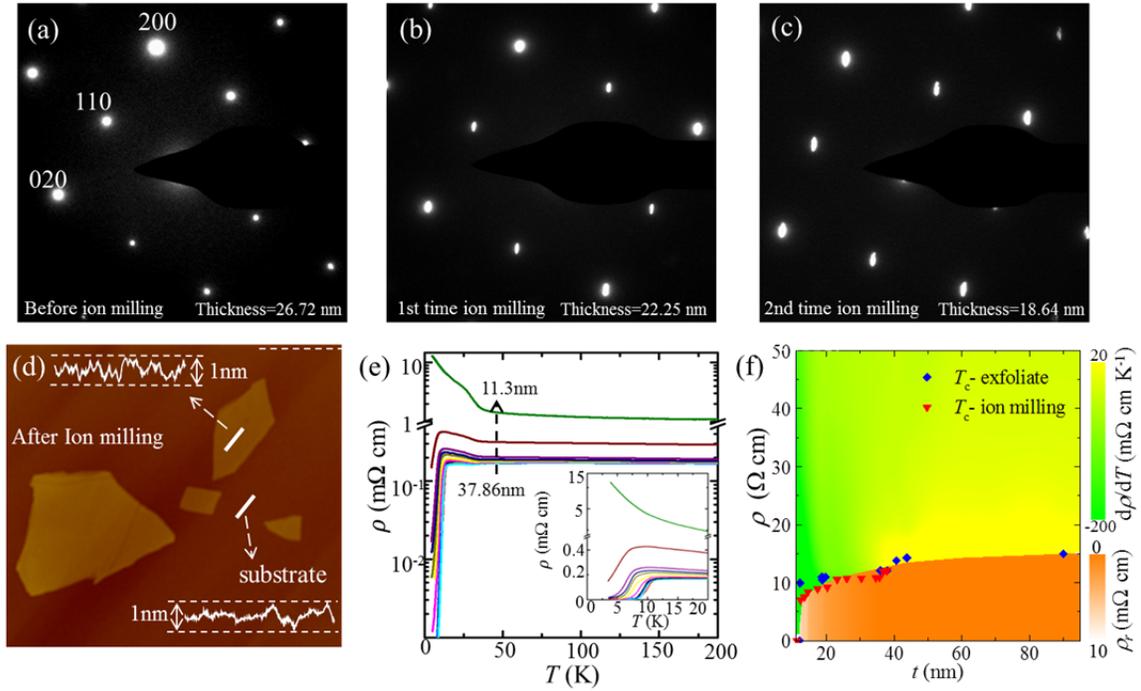

Figure 2. (a)-(c) SAED diffraction patterns of an identical Fe(Te$_{0.5}$Se$_{0.5}$) thin flake before (a) and after two successive (b and c) ion milling thinning. (d) AFM image of the flake after the ion milling. (e) Temperature dependence of resistivity of the Fe(Te$_{0.5}$Se$_{0.5}$) nanodevices obtained using successive ion milling thinning on a 37.86 nm device. (f) Phase diagram of the thickness dependence of superconducting and normal state properties. The contour plots illustrate the derivative of the normal state resistivity $\rho$ and the residual resistivity $\rho_r$ at 2K.

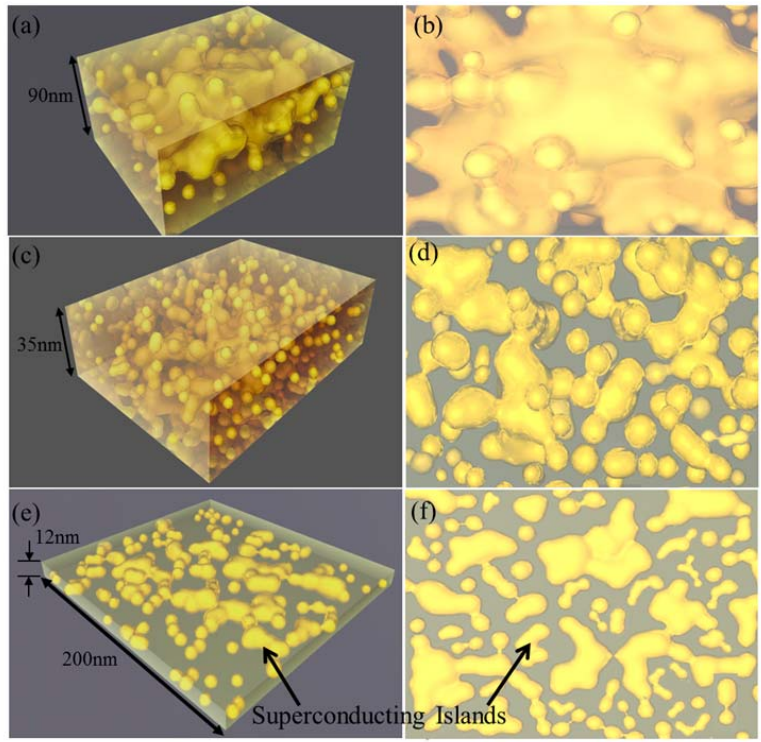

Figure 3. (a), (c), and (e): 3D schematics for the development and connection of the superconducting islands in thick (a), thinner (b), and ultra-thin (e) flakes. (b), (d) and (f): Top view of the 3D schematics in (a), (c), and (e), respectively.

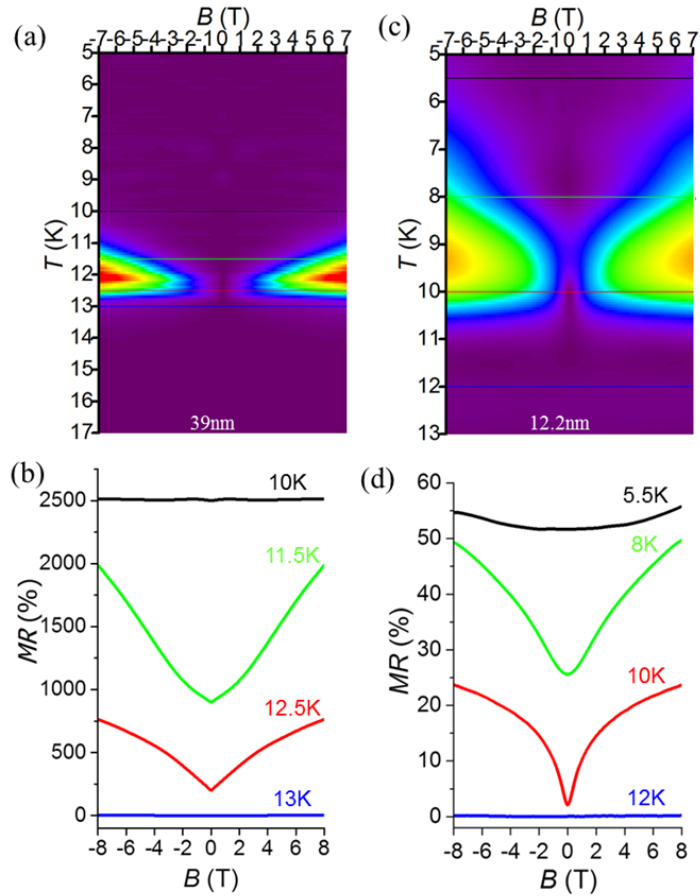

Figure 4. (a) The contour plot of the temperature and magnetic field dependence of magnetoresistivity for a 39 nm flake obtained from microexfoliation. (b) The normalized MR at different temperatures, corresponding to the horizontal cuts in (a). (c) The contour plot of the temperature and magnetic field dependence of magnetoresistivity for a 12.2 nm flake obtained from microexfoliation. (d) The normalized MR at different temperatures, corresponding to the horizontal cuts in (c).